\documentclass{ws-procs9x6}
\def\HTmib#1{\mbox{\boldmath$#1$\unboldmath}}
\def\HTmibsub#1{\mbox{\footnotesize \boldmath$#1$\unboldmath}}
\def\HTxi{\HTmib{x}_{\mbox{\scriptsize i}}}
\def\HTti{t_{\mbox{\scriptsize i}}}
\def\HTxf{\HTmib{x}_{\mbox{\scriptsize f}}}
\def\HTtf{t_{\mbox{\scriptsize f}}}
\def\HTpsif{\psi_{\mbox{\scriptsize f}}}
\def\HTpsii{\psi_{\mbox{\scriptsize i}}}
\def\HTweak{_{\!\mbox{\scriptsize (f;i)}}^{\!\mbox{\scriptsize w}}}
\def\HTket#1{|#1\rangle}
\def\HTbra#1{\langle #1|}
\def\HTDF#1#2{\frac{\partial #1}{\partial #2}}
\def\HTDDF#1#2{\frac{\partial^2 #1}{{\partial #2}^2}}
\begin{document}

\title{CLASSICAL ANALOGUE OF WEAK VALUE IN STOCHASTIC PROCESS}

\author{H. TOMITA}

\address{Research Center of Quantum Computing, Kinki University\\
Higashi-Osaka, 577-8502, Japan\\
E-mail: tomita@alice.math.kindai.ac.jp}

\begin{abstract}
One of the remarkable notions in the recent development 
of quantum physics is the weak value related to weak measurements.
We emulate it as a two-time conditional expectation
in a classical stochastic model.
We use the well known symmetrized form of the master equation, 
which is formally equivalent to the wave equation in quantum
mechanics apart from the fact that wave functions are always real.
The origin of the unusual behaviors of the weak value 
such as the negative probability
and the abnormal enhancement of some expectations becomes clearer
in the present case, where the two-time conditional probability
has no ambiguity of imaginary/complex values.
\end{abstract}


\bodymatter

\section{Introduction}
The weak value is a derived notion of the weak measurement proposed
by Aharonov et al \cite{HT-Aharonov}, which has brought a new 
understanding of quantum observations.
The weak measurement \cite{HT-Measurement} means that it hardly disturbs
the quantum superposed state when it is performed with large uncertainty.
The reason of the strange nature of this quantum measurement is that
the weak value is defined as an expectation over strongly restricted
paths with the condition of the post-selected final state.
\par
Suppose we have started from a pre-selected initial state
at $t=\HTti$.
If we measure some observable $Q$ at $t~(>\HTti)$,
the wave function $\varPsi(t)$
collapses to one of the eigenstates of $Q$.
And we can find only the probability distribution $|\varPsi(t)|^2$,
not the probability amplitude $\varPsi(t)$, 
by repeating the measurement and adopting all of the observed data.
If we discard the main data by restricting the paths to the
post-selected ones only, we can expect to get some informations on
the state $\varPsi(t)$ before collapsing,
because the post-selected paths are described with
the same propagator as the forward evolution of the pre-selected paths
by changing only the sign of the time $t$ according to the time-reversal
symmetry of quantum mechanics.
\par
\medskip
The weak value of an observable $Q$ with a given initial state
$|i\rangle $ at $t=\HTti$ and a final state $|f\rangle $ at $t=\HTtf$
is defined by\cite{HT-Hosoya}
\begin{equation}\label{HT-WV}
\langle Q\rangle \HTweak
=\frac{\langle f|\mbox{e}^{-i(\HTtf -t)H}Q\mbox{e}^{-i(t-\HTti)H}|i\rangle }
{\langle f|\mbox{e}^{-i(\HTtf-\HTti)H}|i\rangle },
~(\HTti\le t\le\HTtf)
\end{equation}
where $H$ is the Hamiltonian and the unit $\hbar=1$ is used.
This quantity is related to a \textit{weak measurement} as follows:
Let us introduce a \textit{meter} to measure the observable $Q$
of the target system at $t=t_0$ by a weak interaction
\[
H_{\mbox{\footnotesize int}}(t)=g\delta(t-t_0)Q\otimes p,
\]
where $p$ is the momentum operator of the probe of the meter and
$g$ is a small coupling constant.
Suppose the initial state of the meter, $\varphi(x)$ in the coordinate
representation of the probe position $x$ has a sufficiently broad
uncertainty $\varDelta$, i.e. a variance $\varDelta^2$.
It can be easily shown that for the restricted paths from $i$ to $f$,
the meter state for $t>t_0$ is given by
\begin{eqnarray}
&&\langle f|\mbox{e}^{-i(\HTtf-t)H}\mbox{e}^{-igQ\otimes p}
\mbox{e}^{-i(t-\HTti)H}|i\rangle \varphi(x)\nonumber\\
&\simeq&
\langle f|\mbox{e}^{-i(\HTtf-\HTti)H}|i\rangle
\varphi\left( x-g\langle Q\rangle \HTweak\right),
\end{eqnarray}
where $p=-i\partial/\partial x$ is used.
Note that the weak value defined by Eq.(\ref{HT-WV}) is complex in general.
Then the shift of the expectation of the probe position $x$ is given by
the real part of the weak value,
$g\mbox{Re}[\langle Q\rangle \HTweak]$,
while the shift of the expectation of the momentum $p$ is found to be
equal to the imaginary part,
$(g/2\varDelta^2)\mbox{Im}[\langle Q\rangle \HTweak]$
by using a Fourier transformation.
\par
\medskip
An early interpretation of the quantity defined by Eq.(\ref{HT-WV}) is
the probability \textit{amplitude}\cite{HT-ABL} which yields
a pre- and post-selected, two-time conditional probability (TTCP).
When it is applied to a projection
operator $|q\rangle\langle q|$ onto one of the eigenstates of
an observable $Q$, one finds
\begin{equation}\label{HT-PAMP}
|\langle|q\rangle\langle q|\rangle\HTweak|^2=
\left|\frac{\langle f|\mbox{e}^{-i(\HTtf -t)H}|q\rangle\langle q|
\mbox{e}^{-i(t-\HTti)H}|i\rangle }
{\langle f|\mbox{e}^{-i(\HTtf-\HTti)H}|i\rangle}\right|^2
=\frac{P(f|q)P(q|i)}{P(f|i)}.
\end{equation}
The last expression is to be shown equal to $P(q|f\cap i)$ in Sec.3
by using Bayes identities. 
(See the footnote c in Sec.3.)
Here $P(*|C)$ is the standard notation for a conditional probability
with a condition $C$ (or a transition probability from
the state $C$ to $*$), e.g.
\[ 
P(f|q)=|\langle f|\mbox{e}^{-i(\HTtf -t)H}|q\rangle|^2,~~\mbox{etc.}
\]
\par
Another interpretation, somewhat formal one, is the TTCP
itself.\cite{HT-Hosoya, HT-HS}
Let us rewrite the usual quantum expectation of
$Q(t)=\mbox{e}^{i(t-\HTti)H}Q\mbox{e}^{-i(t-\HTti)H}$ with respect to
a state $|i\rangle$ in the following identical form,
\begin{eqnarray}\label{HT-QE}
\langle i|Q(t)|i\rangle 
&=&\sum_{f}\langle i|\mbox{e}^{i(\HTtf-\HTti)H}|f\rangle 
\langle f|\mbox{e}^{-i(\HTtf-t)H}Q\mbox{e}^{-i(t-\HTti)H}|i\rangle \nonumber\\
&=&\sum_{f}\langle Q\rangle \HTweak~
|\langle f|\mbox{e}^{-i(\HTtf-\HTti)H}|i\rangle |^2,
\end{eqnarray}
where the last factor of Eq.(\ref{HT-QE})
is $P(f | i)$.
This may yield an interpretation of the weak value as a complex,
raw stochastic variable.
Nevertheless, if it is applied to $|q\rangle \langle q|$ again,
it reads as a conditional probability equation,
\begin{equation}
P(q|i)=|\langle q|\mbox{e}^{-i(t-\HTti)H}|i\rangle |^2=
\sum_{f}\langle|q\rangle \langle q|\rangle\HTweak P(f | i).
\end{equation}
In addition, we have a sum-rule,
\[ \sum_{q}
\langle|q\rangle \langle q|\rangle\HTweak=1, \]
because of the completeness, $\sum_{q}|q\rangle \langle q|=I~
(=\mbox{identity})$.
Therefore, if we remind a type of the Bayes statistics relations,
\footnote{~$P(f\cap q|i)=P(q\cap f\cap i)/P(i)=
P(q|f\cap i)P(f\cap i)/P(i)=P(q|f\cap i)P(f|i)$\\
For the Bayes identity, see the equation $(*)$ in the footnote c in Sec.3.}
\[
P(q | i)=\sum_{f}P(f\cap q | i)=\sum_{f} P(q|f\cap i)P(f | i),
\]
the weak value of the projection operator
$|q\rangle \langle q|$, though it may be \textit{complex}, can be interpreted 
formally as a TTCP itself with a couple of pre- and post-selections,
$i$ and $f$.
Further, the weak value of an operator $Q=\sum_{q}q|q\rangle \langle q|$,
or $A=\sum_{q}a(q)|q\rangle \langle q|$ in general, can be
interpreted as the two-time conditional expectations (TTCE) of them
with respect to this virtual TTCP.
\par
\medskip
Because of this rather fictitious interpretation, the virtual conditional 
probability happens to be \textit{negative}\cite{HT-Feynman} and it causes
an abnormal enhancement of the weak value of some observables 
greater than their inherent norms.\cite{HT-HS} 
These strange behaviors are closely related.
That is, at least if a probability set $\{P(q)\}$ is real,
it can be expected that we have a partial sum satisfying
\begin{equation}
\sum_{P(q)\ge 0}\!\!P(q)=1-\sum_{P(q)<0}\!\!P(q)~>1,
\end{equation}
whenever there exists a negative part.
This is the essential reason of a possibility of the unusual enhancement
of some expectations. \cite{HT-Feynman, HT-Sokolovski} 
\par
\medskip
The purpose of the present work is to emulate these strange behaviors
clearer by using a classical stochastic model,
in which we can avoid the ambiguity of the complex probability
in the above quantum problem \cite{HT-A-B}.
We survey a conventional transformation of the stochastic master equation
to a self-adjoint form in the following section.
A good analogy with the quantum mechanics is found by applying it
to the TTCP. 
This is shown in Sec.3.
An example of the stochastic Ising model which shows an abnormal 
enhancement of the expectations of some quantities with respect to TTCP
is given in Sec.4.
In Sec.5 we discuss an extension of TTCP to a density matrix form
to complete the analogy with the quantum mechanics.
The last section is devoted to brief summary and discussions.

\section{Self-adjoint form of stochastic master equation}
First let us survey the well-known transformation \cite{HT-KMK} 
to a self-adjoint form of the stochastic master equation.
\par
Let \HTmib{x} be a set of stochastic variable(s) described by a time-dependent
conditional probability, 
$P(\HTmib{x},t|\HTxi,\HTti)$ for $t\ge \HTti$, which obeys the following 
stationary, Markovian master equation, i.e. the Chapman-Kolmogorov
\textit{forward} equation,
\begin{eqnarray} \label{HT-Master}
\frac{\partial}{\partial t}P(\HTmib{x},t|\HTxi,\HTti)
&=&-\sum_{\HTmibsub{x'}}W(\HTmib{x}\to\HTmib{x'})P(\HTmib{x},t|\HTxi,\HTti)\nonumber\\
&&~~~~~~~~~~~~~~~~
+\sum_{\HTmibsub{x'}}W(\HTmib{x'}\to\HTmib{x})P(\HTmib{x'},t|\HTxi,\HTti)\nonumber\\
&=&-\sum_{\HTmibsub{x'}}L(\HTmib{x},\HTmib{x'})P(\HTmib{x'},t|\HTxi,\HTti),
\end{eqnarray}
where
\[
L(\HTmib{x},\HTmib{x'})=\delta(\HTmib{x}-\HTmib{x'})\sum_{\HTmibsub{x''}}
W(\HTmib{x}\to\HTmib{x''}) - W(\HTmib{x'}\to\HTmib{x}).
\]
The matrix $L$ has an eigenvalue $\lambda_0=0$ 
corresponding to the steady state,
\[
P_0(\HTmib{x})=\lim_{t-\HTti\to\infty} P(\HTmib{x},t|\HTxi,\HTti).
\]
Let us introduce a \textit{wave function} related to this
forward conditional probability by
\begin{equation}
\psi(\HTmib{x},t|\HTxi,\HTti)=\phi_0(\HTmib{x})^{-1}
P(\HTmib{x},t|\HTxi,\HTti),~~(t\ge \HTti)
\end{equation}
where $\phi_0(\HTmib{x})=P_0(\HTmib{x})^{1/2}$.
This function $\psi$ obeys the forward wave equation,
\begin{equation} \label{HT-Forward}
\HTDF{}{t}\psi(\HTmib{x},t)=-\sum_{\HTmibsub{x'}}
H(\HTmib{x},\HTmib{x'})\psi(\HTmib{x'},t),
\end{equation}
where $H$ is defined by
\begin{equation}
H(\HTmib{x},\HTmib{x'})=\phi_0(\HTmib{x})^{-1}L(\HTmib{x},\HTmib{x'})
\phi_0(\HTmib{x'}).
\end{equation}
For the time being the initial condition $(\HTxi,\HTti)$ in $\psi$ is
abbreviated.
The function $\phi_0(\HTmib{x})$ is an eigenfunction of
Eq.(\ref{HT-Forward}) for $\lambda_0=0$.
\par
The merit of this transformation is that the eigenvalue problem of a given
master equation is simplified, if the matrix $H$ is symmetric, i.e.
\[
H(\HTmib{x},\HTmib{x'})=H(\HTmib{x'},\HTmib{x}).
\]
This situation is widely expected when the detailed balance condition,
i.e. the time-reversal symmetry \cite{HT-Onsager},
\[
P_0(\HTmib{x})W(\HTmib{x}\to\HTmib{x'})
=P_0(\HTmib{x'})W(\HTmib{x'}\to\HTmib{x}),
\]
or equivalently,
\begin{equation} \label{HT-DB}
L(\HTmib{x},\HTmib{x'})P_0(\HTmib{x'})=L(\HTmib{x'},\HTmib{x})P_0(\HTmib{x}),
\end{equation}
is satisfied.
\footnote{
It should be noted that the time reversal symmetry is assumed
in this form of the probability flow and not on the transition probability
itself, the latter being satisfied in quantum mechanics.
This is the reason why we need the above transformation to obtain
a self-adjoint formulation like quantum mechanics.
}
In this case the eigenvalues of $H$ are all real, and non-negative, if
the steady state is stable. Therefore, $\phi_0(\HTmib{x})$ is the
ground state.
\par
A useful example is the Fokker-Planck equation for a single, continuous
stochastic variable $x$,
\begin{equation} \label{HT-FP}
 \HTDF{}{t}P(x,t)=-{\cal L}[x]P(x,t),~
{\cal L}[x]=-\HTDF{}{x}\left(F'(x)+\frac{\epsilon}{2}\HTDF{}{x}\right),
\end{equation}
which describes a one-dimensional Brownian motion in a potential
$F(x)$ with a small diffusion constant $\epsilon$.
By using its steady state solutions,
\[
 P_0(x)\propto \exp\left[ -2F(x)/\epsilon \right]~~\mbox{and}~~
\phi_0(x)\propto \exp\left[ -F(x)/\epsilon \right],
\]
we find the continuous variable version of the above formulations,
\begin{equation}
 {\cal H}[x]=\frac{1}{\epsilon}\left[
-\frac{\epsilon^2}{2}\HTDDF{}{x}+V(x)\right],~~
V(x)=\frac{1}{2}\left[ F'(x)^2-\epsilon F''(x)\right]. 
\end{equation}
Thus the Fokker-Planck equation is transformed into
a self-adjoint form of an imaginary-time Schr\"odinger equation,
\[
 -\epsilon\HTDF{}{t}\psi(x,t)=
\left[-\frac{\epsilon^2}{2}\HTDDF{}{x}+V(x)\right]\psi(x,t),
\]
and its eigenvalue problem results in a familiar one of the
quantum mechanics.
\par
Figure.1 shows an early application \cite{HT-TIK}
to the so-called Kramers escape problem.
The stochastic decay (or escape) rate of the metastable state
in a double-well potential $F(x)$ is given by the first excited
eigenvalue $\lambda_1$ of the corresponding Schr\"odinger
potential $V(x)$.
The first excited state is almost degenerate with the ground
state for a small diffusion constant $\epsilon$.
\begin{figure}[t]
\begin{center}
\psfig{file=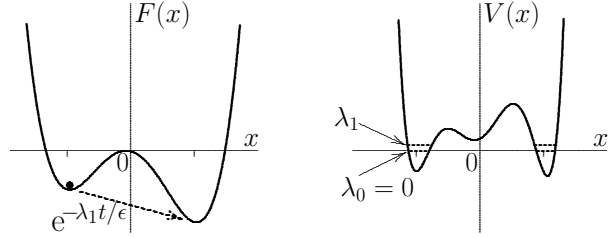,width=8cm}
\caption{\small Stochastic decay process of the metastable state.}
\end{center}
\end{figure}
\section{Two-time conditional probability}
So far the quantum mechanical reformulation merely helps us to simplify
the eigenvalue problem of a given master equation.
None of remarkable quantum mechanical phenomena appears, 
until we are concerned with the TTCP,
\begin{equation}
P(\HTmib{x},t|\HTxf,\HTtf;\HTxi,\HTti),~~\HTti\le t\le \HTtf~.~~~
\mbox{(~$;$ denoting `and', or~}\cap~)
\end{equation}
By using the Markovian property and the well-known relation between
joint and conditional probabilities repeatedly,
\footnote{By using the primitive identity of the Bayes theorem,
\[
P(A|B)P(B)=P(B|A)P(A)=P(A\cap B),~~~~~~~~~~~~(*)
\]
we find in abbreviated notations,
\[ P(x|f\cap i)=\frac{P(f\cap x\cap i)}{P(f\cap i)}
=\frac{P(f|x\cap i)P(x\cap i)}{P(f\cap i)}
=\frac{P(f|x)P(x|i)P(i)}{P(f\cap i)},\]
where the Markovness, i.e. $P(f|x\cap i)=P(f|x)$
is assumed for the time order $\HTtf\ge t\ge \HTti$.
By applying the identity $(*)$ to $P(f|x)$ again, we obtain
a symmetric expression,
\[
P(x|f\cap i)=\frac{P(x|f)P(f)}{P(x)}\frac{P(x|i)P(i)}{P(f\cap i)}
=\frac{1}{R(f,i)}\frac{P(x|f)P(x|i)}{P(x)},
\]
where the first denominator $R(f,i)$ is given by
\[
R(f,i)=\frac{P(f\cap i)}{P(f)P(i)}=\sum_{x}\frac{P(x|f)P(x|i)}{P(x)},
\]
because of the normalization condition,
$\sum_{x}P(x|f\cap i)=1~ \forall f\cap i$.
}
the TTCP can be written in the following form with a pair
of wave functions as
\begin{equation} \label{HT-Twotime}
P(\HTmib{x},t|\HTxf,\HTtf;\HTxi,\HTti)
=\frac{1}{\langle\HTpsif |\HTpsii\rangle }
\overline{\psi}(\HTmib{x},t|\HTxf,\HTtf)\psi(\HTmib{x},t|\HTxi,\HTti),
\end{equation}
where the associated wave function denoted by $\overline{\psi}$ is
related to the so-called ~\textit{posterior}
conditional probability, 
~$\overline{P}(\HTmib{x},t|\HTxf,\HTtf)$ for $t\le \HTtf$,~by
\begin{equation}
\overline{\psi}(\HTmib{x},t|\HTxf,\HTtf)=\phi_0(\HTmib{x})^{-1}
\overline{P}(\HTmib{x},t|\HTxf,\HTtf),
\end{equation}
and obeys the \textit{backward} wave equation,
\begin{equation} \label{HT-Backward}
\HTDF{}{t}\overline{\psi}(\HTmib{x},t) = \sum_{\HTmibsub{x'}}
H^{\dagger}(\HTmib{x},\HTmib{x'})\overline{\psi}(\HTmib{x'},t).
\end{equation}
Here $H^{\dagger}$ is the hermite conjugate of $H$, i.e. 
the transposed matrix in the present case. 
The eigensystem is common with the forward equation Eq.(\ref{HT-Forward}),
when $H$ is hermitian, i.e. real and symmetric as has been assumed here.\par
The denominator in Eq.(\ref{HT-Twotime}) is the weight of overlap
between the two wave functions defined by an inner product,
\begin{equation} \label{HT-Inner}
\langle\HTpsif|\HTpsii\rangle 
=\sum_{\HTmibsub{x}}\overline{\psi}(\HTmib{x},t|\HTxf,\HTtf)
\psi(\HTmib{x},t|\HTxi,\HTti).
\end{equation}
Of course this quantity is real, while the corresponding quantity
in the quantum mechanics is complex in general.
\par
Let us define the ket- and the bra-vectors by
\begin{equation}\label{braket}
\HTket{\HTpsii(t)}=\{\psi(\HTmib{x},t|\HTxi,\HTti)\}^{T}~~\mbox{and}~~
\HTbra{\HTpsif(t)}=\{\overline{\psi}(\HTmib{x},t|\HTxf,\HTtf)\}.
\end{equation}
Then the wave equations Eqs.(\ref{HT-Forward}) and
(\ref{HT-Backward}) by assuming $H^{\dagger}=H$ 
are rewritten in the quantum mechanical form as
\begin{equation}
\HTDF{}{t}\HTket{\HTpsii(t)}=-H\HTket{\HTpsii(t)}~~\mbox{and}~~
\HTDF{}{t}\HTbra{\HTpsif(t)}=\HTbra{\HTpsif(t)}H,
\end{equation}
respectively.
Henceforth, $H$ is called the Hamitonian.
\par
By using this pair of the Schr\"{o}dinger equations it is shown
that the overlap integral, or the inner product
$\langle\HTpsif|\HTpsii\rangle $ given by Eq.(\ref{HT-Inner}) does
not depend on the current time $t$, i.e.
\[
\HTDF{}{t}\langle\HTpsif|\HTpsii\rangle
=\langle\HTpsif(t)|H|\HTpsii(t)\rangle 
-\langle\HTpsif(t)|H|\HTpsii(t)\rangle =0.
\]
It should be noted that the present wave function $\psi$ 
satisfies a conservation law only in this meaning
Eq.(\ref{HT-Inner}) coupled with its adjoint $\overline{\psi}$.
In addition, it can be shown that this overlap integral has the following
properties in the respective limits;
\begin{equation} \label{HT-Limit0}
\begin{array}{l}
(\mbox{i})~\displaystyle{\lim_{\HTtf-\HTti\to\infty}}
\langle\HTpsif|\HTpsii\rangle =1,\\
(\mbox{ii})~\displaystyle{\lim_{\HTtf-\HTti\to 0}}\langle\HTpsif|\HTpsii\rangle 
=[\phi_0(\HTxf)\phi_0(\HTxi)]^{-1} \delta(\HTxf-\HTxi).
\end{array}
\end{equation}
\par
Note that the TTCE (two-time conditional expectation) of a physical
quantity $Q$ with respect to TTCP defined by
\begin{equation} \label{HT-TTCE}
\langle Q\rangle \HTweak
=\sum_{\HTmibsub{x}} Q(\HTmib{x})P(\HTmib{x},t|\HTxf,\HTtf;\HTxi,\HTti)
=\frac{\langle \HTpsif(t)|Q|\HTpsii(t)\rangle }{\langle \HTpsif|\HTpsii\rangle },
\end{equation}
has just the analogous form of the weak value in the quantum mechanics.
\par
\medskip
Thus the TTCP is a nonlinear quantity composed 
of a product of a pair of the forward
and the backward wave functions, and cannot be described by a closed,
linear evolution equation.
Then it happens that the principle of the probability superposition
is violated and the interference of wave functions may occur.
However, its example is omitted here because none of 
remarkable phenomena from this view point has been found, yet.
The reason may be that the wave functions are always \textit{real}
and positive in the present case.
Therefore, let us discuss only the weak value in the rest.
\section{Stochastic model of classical Ising spins}
An example is a pair of the classical Ising spin $\sigma=\pm 1$
having an exchange interaction, 
\[
E(\HTmib{x})=-J\sigma_1\sigma_2,
\]
where $\HTmib{x}=(\sigma_1,\sigma_2)$.
Let us number the stochastic variable $\HTmib{x}$ in the order, 
$(1,1)$,~$(1,-1)$,~$(-1,1)$,~$(-1,-1)$ and choose the following 
transition matrices,
\begin{equation}
 W=\left(
\begin{array}{cccc}
0&1&1&0\\
p^2&0&0&p^2\\
p^2&0&0&p^2\\
0&1&1&0
\end{array}
\right)~~\mbox{or}~~
L=\left(
\begin{array}{cccc}
2p^2&-1&-1&0\\
-p^2&2&0&-p^2\\
-p^2&0&2&-p^2\\
0&-1&-1&2p^2
\end{array}
\right),
\end{equation}
where $p=\mbox{e}^{-\beta J},~\beta=1/k_{\mbox{\scriptsize B}}T$.
Evidently this transition matrix $W$ satisfies the detailed balance condition,
\[
\mbox{e}^{-\beta E(\HTmib{x})}W(\HTmib{x}\to\HTmib{x'})
=\mbox{e}^{-\beta E(\HTmib{x'})}W(\HTmib{x'}\to\HTmib{x}),
\]
at the steady state, i.e. the thermal equilibrium of a temperature $T$.
With the use of the equilibrium distribution function,
\[
P_0(\HTmib{x})=\frac{1}{2(1+p^2)}~(1,~p^2,~p^2,~1)
~~\mbox{and}~~
\phi_0(\HTmib{x})=\frac{1}{\sqrt{2(1+p^2)}}~(1,~p,~p,~1),
\]
we find the corresponding hermitian Hamiltonian,
\begin{eqnarray}
H&=&\left(
\begin{array}{cccc}
2p^2&-p&-p&0\\
-p&2&0&-p\\
-p&0&2&-p\\
0&-p&-p&2p^2
\end{array}
\right)
\nonumber\\
&=&(1+p^2)~\sigma_0\otimes\sigma_0
-(1-p^2)~\sigma_z\otimes\sigma_z-p~
(\sigma_0\otimes\sigma_x+\sigma_x\otimes\sigma_0),
\end{eqnarray}
where $\sigma_x$ and $\sigma_z$ are the usual Pauli matrices and
$\sigma_0$ denotes the two dimensional unit matrix $I_2$.
This is the Hamiltonian of a pair of \textit{quantum}
Ising spins with an exchange interaction in a transverse magnetic field.
\par
The eigenvalues and the eigenstates of this Hamiltonian $H$,
\smallskip
\begin{equation}\label{HT-Eigensystem}
\left\{~
\begin{array}{l}
\lambda_0=0,~\lambda_1=2p^2,~\lambda_2=2,~\lambda_3=2(1+p^2),\\
\\
\HTket{0}=
\displaystyle{\frac{1}{\sqrt{2(1+p^2)}}}\left[~\HTket{\!\uparrow\uparrow}~
+~p~\HTket{\!\uparrow\downarrow}+~p~\HTket{\!\downarrow\uparrow}~
+~\HTket{\!\downarrow\downarrow}~\right],
\\
\HTket{1}=\displaystyle{\frac{1}{\sqrt{2}}}
\left[~\HTket{\!\uparrow\uparrow}~-~\HTket{\!\downarrow\downarrow}~
\right],\\
\HTket{2}=\displaystyle{\frac{1}{\sqrt{2}}}
\left[~\HTket{\!\uparrow\downarrow}~-~\HTket{\!\downarrow\uparrow}~
\right],\\
\HTket{3}=\displaystyle{\frac{1}{\sqrt{2(1+p^2)}}}
\left[~p~\HTket{\!\uparrow\uparrow}~
-~\HTket{\!\uparrow\downarrow}-~\HTket{\!\downarrow\uparrow}~
+~p~\HTket{\!\downarrow\downarrow}~\right],
\end{array}
\right.
\end{equation}
\smallskip
can be easily obtained, where $\HTket{0}=\HTket{\phi_0}$, the ground state.
Here the familiar notations $\uparrow,\downarrow$ are used for
$\sigma=\pm 1$.
Note that the first excited state is almost degenerate with the ground
state for a small transition probability $p^2$.
\par
By using this eigensystem we can calculate the state vectors,
$\HTket{\HTpsii(t)}$
and $\HTbra{\HTpsif(t)}$ for arbitrary initial and final states in just the same
manner of the elementary quantum mechanics except for the fact that
the time $t$ is imaginary.\par
Strange behaviors can be expected only when the paths from $i$ to $f$
are very rare cases, because the post-selection causes little effect
when the paths are dominant ones.
Then let us consider the case where the initial and the final states
differ from each other. Let
\[
\HTxi=\uparrow\uparrow~\mbox{at}~t=0~~\mbox{and}~~\HTxf
=\downarrow\downarrow~\mbox{at}~t=\HTtf,
\]
that is,
\[
P(\HTmib{x},0)=(1,0,0,0)~~\mbox{and}~~\overline{P}(\HTmib{x},\HTtf)=(0,0,0,1),
\]
or equivalently,
\[
\HTket{\HTpsii(0)}=\sqrt{2(1+p^2)}~\HTket{\!\uparrow\uparrow}~~\mbox{and}~~
\HTbra{\HTpsif(\HTtf)}=\sqrt{2(1+p^2)}~\HTbra{\downarrow\downarrow\!}.
\]
By using the eigenvector expansion we obtain,
\begin{equation} \label{HT-InitialWF}
\begin{array}{lcl}
\HTket{\HTpsii(t)}&=&\HTket{0}+
\sqrt{1+p^2}~\mbox{e}^{-\lambda_1 t}~\HTket{1}
+p~\mbox{e}^{-\lambda_3 t}~\HTket{3},\\
\\
\HTbra{\HTpsif(t)}&=&\HTbra{0}
-\sqrt{1+p^2}~\mbox{e}^{-\lambda_1(\HTtf-t)}\HTbra{1}
+p~\mbox{e}^{-\lambda_3(\HTtf-t)}\HTbra{3},
\end{array}
\end{equation}
and
\begin{equation} \label{HT-Overlap}
\langle\HTpsif|\HTpsii\rangle =1-(1+p^2)~\mbox{e}^{-\lambda_1\HTtf}
+p^2~\mbox{e}^{-\lambda_3\HTtf}~(>0).
\end{equation}
\par
The TTCP is shown in Figure.2.
This result itself is very natural and well-expected,
all probabilities being always non-negative.
\begin{figure}[t]
\vskip 0.3cm
\begin{center}
\psfig{file=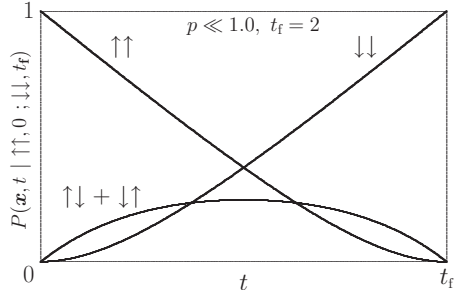,width=6cm}
\caption{\small Two-time conditional probability}
\end{center}
\end{figure}
\par
\medskip
A strange behavior appears when we use the basis
$\{\HTket{k},~k=0,1,2,3\}$, the eigenstates of the Hamiltonian
$H$ instead of the spin states $\{|\HTmib{x}\rangle 
=|\sigma_1\sigma_2\rangle \}$.
We can calculate the virtual probability, i.e. the TTCE of the projection
operator $|k\rangle \langle k|$ onto each eigenstate $|k\rangle $
in the same manner.
The result is given by
\smallskip
\begin{equation} \label{HT-Projection}
\begin{array}{cclcl}
P(0,t)&=&\displaystyle{\frac{\langle\HTpsif(t)|0\rangle
\langle 0|\HTpsii(t)\rangle }
{\langle\HTpsif|\HTpsii\rangle }}
&=&\displaystyle{\frac{1}{\langle\HTpsif|\HTpsii\rangle }}~,\\

P(1,t)&=&
\displaystyle{\frac{\langle\HTpsif(t)|1\rangle \langle 1|\HTpsii(t)\rangle }
{\langle\HTpsif|\HTpsii\rangle }}
&=&-\displaystyle{
\frac{(1+p^2)\mbox{e}^{-\lambda_1\HTtf}}{\langle\HTpsif|\HTpsii\rangle }
}~~(<0~)~,\\

P(2,t)&=&\displaystyle{
\frac{\langle \HTpsif(t)|2\rangle \langle  2|\HTpsii(t)\rangle }
{\langle \HTpsif|\HTpsii\rangle }}
&=&0~,\\
P(3,t)&=&\displaystyle{
\frac{\langle \HTpsif(t)|3\rangle \langle  3|\HTpsii(t)\rangle }
{\langle \HTpsif|\HTpsii\rangle }}
&=&\displaystyle{
\frac{p^2\mbox{e}^{-\lambda_3\HTtf}}{\langle \HTpsif|\HTpsii\rangle }}~.
\end{array}
\end{equation}
\smallskip
The fictitious negative probability is found in $P(1,t)$.
Of course the completeness of the probability,
$\sum_{k=0}^3 P(k,t)=1,$
is satisfied evidently because of Eq.(\ref{HT-Overlap}).
\par
\medskip
This negativity is precisely expected from the signs of the expansion
coefficients of the eigenvectors in the right side
of Eq.(\ref{HT-Eigensystem}).
Some of inner products $\langle  \HTmib{x}|k\rangle $ between two basis systems 
$\{|\HTmib{x}\rangle =|\sigma_1\sigma_2\rangle \}$ and $\{|k\rangle \}$
are found to be negative.
Then some part of the virtual TTCP happens to be negative,
when the initial and the final states differ from each other.
\par
On the contrary, when the both states are
the same, this situation cannot be expected, because the negative
inner products, if any, would be squared.
For example, when we select as
\[ \HTxi=\HTxf=\uparrow\uparrow~\mbox{at}~t=0~\mbox{and}~t=\HTtf, \]
we find the corresponding virtual probabilities all positive, i.e.
\footnote{
Note that the bra- and the ket-vectors
in these expressions denote
\[ |\HTpsii(t)\rangle =\mbox{e}^{-tH}|\HTpsii\rangle ~~\mbox{and}~~
\langle  \HTpsii(t)|=\langle \HTpsii|\mbox{e}^{-(\HTtf-t)H}, \]
from the present definitions Eq.(\ref{braket}) of them,
and the overlap integral 
$\langle \HTpsii|\HTpsii\rangle $ in the denominators is better to be written 
explicitly as
$\langle \HTpsii(t)|\HTpsii(t)\rangle$, or 
$\langle \HTpsii|\mbox{e}^{-\HTtf H}|\HTpsii\rangle $ to avoid
a confusion,
as if $\langle \HTpsii|\HTpsii\rangle =1$ in the usual quantum mechanical notation.
}
\medskip
\begin{equation}
\begin{array}{cclcl}
P(0,t)&=&\displaystyle{
\frac{\langle \HTpsii(t)|0\rangle\langle  0|\HTpsii(t)\rangle }
{\langle \HTpsii|\HTpsii\rangle }
}
&=&\displaystyle{\frac{1}{\langle \HTpsii|\HTpsii\rangle }}~,\\

P(1,t)&=&\displaystyle{
\frac{\langle \HTpsii(t)|1\rangle \langle  1|\HTpsii(t)\rangle }
{\langle \HTpsii|\HTpsii\rangle }
}
&=&\displaystyle{
\frac{(1+p^2)\mbox{e}^{-\lambda_1\HTtf}}{\langle \HTpsii|\HTpsii\rangle }
}~,\\

P(2,t)&=&\displaystyle{
\frac{\langle \HTpsii(t)|2\rangle \langle  2|\HTpsii(t)\rangle }
{\langle \HTpsii|\HTpsii\rangle }
}
&=&0~,\\
P(3,t)&=&\displaystyle{
\frac{\langle \HTpsii(t)|3\rangle \langle  3|\HTpsii(t)\rangle }
{\langle \HTpsii|\HTpsii\rangle }
}
&=&\displaystyle{
\frac{p^2\mbox{e}^{-\lambda_3\HTtf}}{\langle \HTpsii|\HTpsii\rangle }
}~,
\end{array}
\end{equation}
where
\[
\langle \HTpsii|\HTpsii\rangle =1+(1+p^2)~\mbox{e}^{-\lambda_1\HTtf}
+p^2~\mbox{e}^{-\lambda_3\HTtf}.
\]
\par
\medskip
This is a general conclusion for two different bases
$\{\HTmib{e}_i\}$ and $\{\HTmib{e}_j'\}$ of any \textit{real}
vector space, because
at least one of the inner products, $\{\HTmib{e}_i\cdot\HTmib{e}_j'\}$
must be negative.
\par
\medskip
That is, the negative probability can be expected at least when
\begin{itemize}
\item[(1)]{the initial and the final states differ from each other,}
\item[(2)]{and the orthogonal basis of the intermediate projection
differs from the basis of the initial and the final selections,
so that one of the inner products is negative.}
\end{itemize}
The same situation may occur in the quantum system between 
different sets of eigenvectors of \textit{non-commutative} observables, say,
$P, Q$.
When we select the initial and the 
final states as different eigenstates of $P$,
the virtual TTCP, i.e. the weak value of the projection operator
$|q\rangle\langle q|$ onto some of the eigenstates of $Q$
can be negative.
This setting is sufficient for the condition
\cite{HT-HS} to find the strange weak value.
\par
\medskip
\begin{figure}[t]
\vskip 0.5cm
\begin{center}
\psfig{file=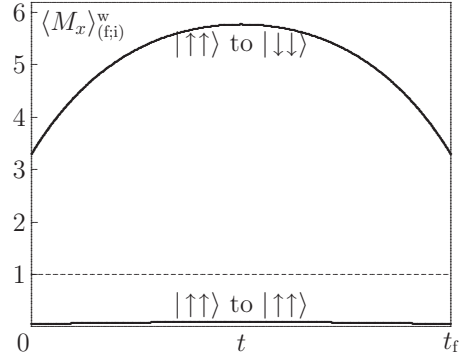,width=6cm}
\caption{\small Abnormal and normal TTCE of the
transverse magnetization $M_x$ for the transition probability
$p=0.2$ and $p^2\HTtf=0.01$. The indicators $|\!\!\uparrow\uparrow\rangle$
and $|\!\!\downarrow\downarrow\rangle$ denote
the initial and the final states of the respective TTCP.}
\vskip -0.5cm
\end{center}
\end{figure}
A strange behavior related to this negative probability
is the abnormal enhancement
of some observables as is stated in Sec.1.
An example is shown in Fig.3 for a quantity, say, the 
\textit{transverse magnetization},
\begin{equation} \label{HT-Nondiagonal}
M_x=\frac{1}{2}(\sigma_x\otimes\sigma_0+\sigma_0\otimes\sigma_x).
\end{equation}
An abnormal behavior
\begin{eqnarray}
\langle  M_x\rangle \HTweak
&=&\frac{1}{\langle \HTpsif|\HTpsii\rangle }
\left[\displaystyle{
\frac{2p}{1+p^2}\left(1-p^2\mbox{e}^{-\lambda_3\HTtf}\right)
-\frac{1-p^2}{1+p^2}\left(\mbox{e}^{-\lambda_3 t}
+\mbox{e}^{-\lambda_3(\HTtf-t)}\right)}\right]\nonumber\\
&>&1,
\end{eqnarray}
is found for sufficiently small $p$ and $\HTtf$.
Note that the natural norm of $M_x$ must be less than 1,
because the eigenvalue spectrum of $M_x$ is $\{-1, 0, 0, 1\}$.
When the transition rate is very small, 
i.e. $p^2 \HTtf\ll 1$, we find
\[
\langle  M_x\rangle \HTweak\gg 1.
\]
A plain reason of this singular behavior is that the overlap integral
$\langle \HTpsif|\HTpsii\rangle $ in the denominator may be expected
to be very small owing to (ii) of Eq.(\ref{HT-Limit0}),
whenever the initial and the final states differ from each other,
i.e. $\HTxi\ne\HTxf$.
This means that to reach $\HTxf=(\downarrow\downarrow)$ starting
from $\HTxi=(\uparrow\uparrow)$ in a given time occurs scarcely
and is far from the main flow of the conditional probability.
On the contrary none of such strange behaviors are found
when $\HTxi=\HTxf$, e.g. $\HTxi=\HTxf=(\uparrow\uparrow)$.
The result for the latter case for the same parameters as the upper
abnormal case is shown by the lower curve in Fig.3,
its maximum being $\sim 0.09$ at $t=\HTtf/2$
and minimum $\sim 0.05$ at $t=0$ and $\HTtf$.
\par
\medskip
In Fig.4 the TTCE of another
quantity $A=\sigma_x\otimes\sigma_x$ having a spectrum $\{-1,-1,1,1\}$
are shown also.
Note that $A$ is commutative with $H$ and is a conserved quantity.
Then the horizontal axis in this figure shows a parameter 
of the transition probability instead of the current time itself.
\begin{figure}[t]
\vskip 0.5cm
\begin{center}
\psfig{file=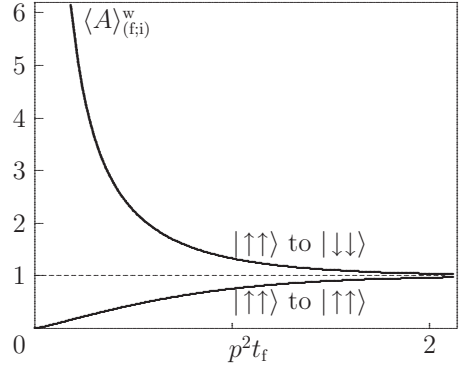,width=6cm}
\caption{\small Abnormal and normal TTCE of $A=\sigma_x\otimes
\sigma_x$ for $p=0.2$.}
\vskip -0.5cm
\end{center}
\end{figure}
\par
\noindent
\textbf{(Single spin model)}~~The previous model having an energy barrier
is aimed at realizing the rare paths from $i$ to $f$ 
in order to find out a strange weak value easily.
After the symposium, however, the author noticed that
a single, free Ising spin system provides a simpler example.
\par
\medskip
An Ising spin flips up-to-down and down-to-up randomly.
A transition probability may be defined by
\begin{equation}
W=\left(\begin{array}{cc}0&~p\\p&~0\end{array}\right)~~
\mbox{and}~~
L=\left(\begin{array}{cc}p&-p\\-p&p\end{array}\right),
\end{equation}
whose equilibrium state is
\[ P_0(\HTmib{x})=\left(\frac{1}{2},\frac{1}{2}\right)~~\mbox{or}~~
\phi_0(\HTmib{x})=\left(\frac{1}{\sqrt{2}},\frac{1}{\sqrt{2}}\right). \]
Then, the Hamiltonian for this classical stochastic model is given by
\footnote{
A quantum free spin system has been used often to demonstrate
the weak value in quantum mechanics, but it differs from
the present system which has a transverse magnetic field $p$
as shown by this Hamiltonian.
}
\begin{equation}\label{HT-FREE}
H=\left(\begin{array}{cc}p&-p\\-p&p\end{array}\right)
=p(\sigma_0-\sigma_x).
\end{equation}
Another eigenstate, i.e. the excited state is
\[ \lambda_1=2p~~\mbox{and}~~\phi_1(\HTmib{x})=\left(\frac{1}{\sqrt{2}},-\frac{1}{\sqrt{2}}\right),
\]
that is, the two eigenstates of $H$ are those of $\sigma_x$ itself, i.e.
the Hadamard states,
\begin{equation}
|0\rangle=\frac{1}{\sqrt{2}}(|\!\uparrow\rangle+|\!\downarrow\rangle)~~\mbox{and}~~
|1\rangle=\frac{1}{\sqrt{2}}(|\!\uparrow\rangle-|\!\downarrow\rangle).
\end{equation}
\par
Let us select the initial and the final states as
\[
\HTxi=\uparrow~~\mbox{at}~~t=0,~~\HTxf=\downarrow~~\mbox{at}~~t=\HTtf.
\]
Then we find
\begin{equation}
|\HTpsii(t)\rangle=|0\rangle+\mbox{e}^{-2pt}|1\rangle,
~~\langle\HTpsif(t)|=\langle 0|-\mbox{e}^{-2p(\HTtf-t)}\langle 1|,
\end{equation}
and
\begin{equation}
\langle\HTpsif|\HTpsii\rangle=1-\mbox{e}^{-2p\HTtf},
\end{equation}
by using the eigenvector expansion, where the negative expansion
coefficient appears.
After the same procedures as the previous model 
we obtain an extraordinary TTCP,
\begin{equation}
\begin{array}{ccccrc}
P(0,t)&=&
\displaystyle{
\frac{\langle\HTpsif(t)|0\rangle\langle 0|\HTpsii(t)\rangle}
{\langle\HTpsif|\HTpsii\rangle}
}
&=&\displaystyle{\frac{1}{1-\mbox{e}^{-2p\HTtf}}} & >1,\\
P(1,t)&=&
\displaystyle{
\frac{\langle\HTpsif(t)|1\rangle\langle 1|\HTpsii(t)\rangle}
{\langle\HTpsif|\HTpsii\rangle}}
&=&-\displaystyle{\frac{\mbox{e}^{-2p\HTtf}}{1-\mbox{e}^{-2p\HTtf}}} & <0,
\end{array}
\end{equation}
and a strange weak value,
\begin{equation}
\langle\sigma_x\rangle\HTweak=\coth p\HTtf >1,
\end{equation}
again.
\par
On the contrary, when $\HTxi=\HTxf=\uparrow$,
we find an ordinary TTCP,
\begin{equation}
\begin{array}{cccc}
P(0,t)&=&\displaystyle{\frac{1}{1+\mbox{e}^{-2p\HTtf}}}&>0,\\
P(1,t)&=&\displaystyle{\frac{\mbox{e}^{-2p\HTtf}}{1+\mbox{e}^{-2p\HTtf}}}&>0,
\end{array}
\end{equation}
and
\[ \langle\sigma_x\rangle\HTweak=\tanh p\HTtf <1.\]
\par
Thus the origin of the negative probability and the strange weak value
is more obvious in this simplest system.
\section{Extension of TTCP to a density matrix}
It should be noted that the physical quantities $M_x$ and $A$
in the previous section are
\textit{non-diagonal} in the spin-state representation and have no 
corresponding quantities in the classical Ising spin system.
They are related to the transition rate of the stochastic Ising spin.
In order to calculate the expectations of such non-diagonal quantities
we need an extension of the TTCP to the two-time conditional 
density matrix defined by
\begin{eqnarray} \label{HT-Density}
\rho\HTweak(t)
&=&\frac{1}{\langle \HTpsif|\HTpsii\rangle }~
|\HTpsii(t)\rangle \langle \HTpsif(t)|\nonumber\\
&=&\frac{1}{\langle \HTpsif|\HTpsii\rangle }
\sum_{\HTmibsub{x},\HTmibsub{x'}}
\overline{\psi}(\HTmib{x'},t|\HTxf,\HTtf)\psi(\HTmib{x},t|\HTxi,0)~
\HTket{\HTmib{x}}\HTbra{\HTmib{x'}}.
\end{eqnarray}
From the definition Eq.(\ref{HT-Inner}) of the overlap integral
$\langle \HTpsif|\HTpsii\rangle $, it is evident that
\[
\mbox{Tr}~\rho\HTweak(t)=
\frac{1}{\langle \HTpsif|\HTpsii\rangle }
\sum_{\HTmibsub{x}}
\overline{\psi}(\HTmib{x},t|\HTxf,\HTtf)\psi(\HTmib{x},t|\HTxi,0)
=1.
\]
It should be noted, however, that the diagonal elements of this
density matrix are not always positive as is shown by Eq.(\ref{HT-Projection})
in Sec.4, when it is diagonalized by using the basis
$\{|k\rangle ,k=0,1,2,3\}$, the eigenstates of the Hamiltonian $H$.
\par
With the use of this density matrix the definition Eq.(\ref{HT-TTCE})
of the TTCE is extended as
\[
\langle  Q\rangle \HTweak=
\mbox{Tr}~\rho\HTweak Q.
\]
Of course this definition of the TTCE results in the classical one,
if $Q$ is a diagonal quantity.\par
The notion of this density matrix has not been used in the conventional
classical stochastic process.
It should be emphasized, however, that this quantity is within
a scheme of the classical stochastic process itself,
because the wave functions,
$\psi$ and $\overline{\psi}$ in Eq.(\ref{HT-Density}) are related to the
forward and the posterior, classical conditional probabilities,
respectively.
In addition, we have an alternative expression for $\overline{\psi}$,
\begin{equation} \label{HT-Inverse}
\overline{\psi}(\HTmib{x'},t|\HTxf,\HTtf)
=\psi(\HTmib{x'},\HTtf|\HTxf,t)~
\left(=\phi_0(\HTmib{x'})^{-1}P(\HTmib{x'},\HTtf-t|\HTxf,0)\right),
\end{equation}
or equivalently,
\begin{eqnarray}
\overline{P}(\HTmib{x'},t|\HTxf,\HTtf)P_0(\HTxf)&=&P(\HTxf,\HTtf|\HTmib{x'},t)P_0(\HTmib{x'})
\nonumber\\
&=&P(\HTmib{x'},\HTtf|\HTxf,t)P_0(\HTxf),
\end{eqnarray}
for $t\le\HTtf$ due to the time-reversal symmetry corresponding to the 
detailed balance.
Then the density matrix Eq.(\ref{HT-Density}) can be written as
\begin{equation}
\rho\HTweak(t)
=\frac{1}{\langle \HTpsif|\HTpsii\rangle }
\sum_{\HTmibsub{x},\HTmibsub{x'}}
\displaystyle{
\frac
{P(\HTmib{x'},\HTtf -t|\HTxf,0)P(\HTmib{x},t|\HTxi,0)}
{\phi_0(\HTmib{x'})\phi_0(\HTmib{x})}
}
\HTket{\HTmib{x}}\HTbra{\HTmib{x'}},
\end{equation}
while the overlap integral can be re-defined by
\begin{equation}
\langle \HTpsif|\HTpsii\rangle=
\sum_{\HTmibsub{x}}
\displaystyle{
\frac
{P(\HTmib{x},\HTtf -t|\HTxf,0)P(\HTmib{x},t|\HTxi,0)}
{P_0(\HTmib{x})}
}.
\end{equation}
This fact means that we can define the TTCP and the corresponding
density matrix with only a pair of the usual, forward conditional
probabilities for two individual initial states, $\HTxi$ and $\HTxf$.
We need no data discarding due to the post-selection.
\section{Summary and discussions}
Except for the facts that the time is imaginary and the wave function is 
always real and positive, the classical stochastic process can be 
described in an analogous form of the quantum mechanics, if we use 
the TTCP.
For example, the abnormal behaviors of the weak value in the quantum
mechanics are emulated.
The TTCP and its TTCE, i.e. the weak values are always
real in the present classical case.
Therefore, the origin of such abnormal behaviors is clearer than the 
quantum mechanical case where complex quantities appear.
\par
In addition, if we have not the explicit solution of the eigenvalue problem,
we may calculate the weak value at least with use of a Monte-Carlo
simulation which is often used to investigate the stochastic model.
In performing a simulation it should be noted that
we can calculate the TTCP and its TTCE with two usual, forward conditional
probabilities for respective \textit{initial} conditions, 
the pre-selected and the post-selected ones,
when the detailed balance condition is satisfied.
\par
The importance of the weak value in the quantum mechanics
is that it is related to the new notion of the
weak measurement without disturbing the quantum state.
An analogous notion of the latter in the classical stochastic process,
if any, has not been found yet.
\section*{Acknowledgment}
This work is supported by Open Research Center Project
for Private Universities: Matching fund subsidy from
MEXT of Japan.


\end{document}